\documentclass[preprint,showpacs,amsmath,amssymb]{revtex4-1}
\usepackage{amsmath}
\usepackage{amssymb}
\usepackage{graphicx}
\usepackage{dcolumn} 
\usepackage{bm} 

\begin{document}

\title{Non-reciprocal light scattering by lattice of magnetic vortices}

\author{O.G.Udalov}
\email{udalov@ipmras.ru}
\author{M.V.Sapozhnikov}
\email{msap@ipmras.ru}
\author{E.A.Karashtin}
\author{B.A.Gribkov}
\author{S.A.Gusev}
\author{E.V.Skorohodov}
\author{V.V.Rogov}
\author{A.Yu.Klimov}
\author{A.A.Fraerman}

\affiliation{Institute for Physics of Microstructures RAS, Nizhny Novgorod 603950, GSP-105, Russia}

\date{\today}

\begin{abstract}
We report on experimental study of optical properties of two-dimensional square lattice of triangle Co and CoFe nanoparticles with a vortex magnetization distribution. We demonstrate that intensity of light scattered in  diffraction maxima depends on the vorticity of the particles magnetization and it can be manipulated by applying an external magnetic field. The experimental results can be understood in terms of phenomenological theory.
\end{abstract}


\maketitle

Considerable achievements in micro- and nano- technologies open the new possibilities in fabrication of artificial nanomaterials (metamaterials) with novel interesting optical properties. Planar metallic structures attracts attention due to a lot of interesting effects appears in them. Namely, there are enhanced magneto-optical effects \cite{Belo11,Gitt07,Sap11}, extraordinary light transmission through the subwavelength holes \cite{Genet07} and effective generation of the second harmonic \cite{Anc03,Lamb97,Murzina2009}. The special attention is paid to planar structures consisting of chiral elements \cite{Papa03,Vall03,Schw03,Zhang06,Pros05,Bass88}. Such structures are ordinarily a two-dimensional regular lattice of non magnetic particles that do not possess the reflection symmetry (in the plane perpendicular to the sample surface) and can be characterized by a {\it pseudovector (axial vector)}. This leads to the phenomenon of asymmetric polarization conversion \cite{Pros06}. In the present paper we explore magnetic planar chiral structure. As it will be shown below such structures can be characterized by a {\it polar vector} that change its sign under the time reversal. This causes non-reciprocal optical effects \cite{Groen62,Krich98,Brown63}, whose cannot exist in non magnetic structures.

In our work we investigate regular lattices of the particles with the vortex magnetization distribution. In contrast to non magnetic planar chiral structure spatial inversion symmetry is broken in the vortex particle due to the non-trivial magnetization distribution. The vorticity direction can be manipulated by the tip of magnetic-force microscope \cite{Mir07}, by uniform external magnetic field \cite{Yak10} or by applying an electric current \cite{Yang11}. Making the magnetic particles of a triangular shape one can get all the particles to have the same vorticity and thus the same planar chirality by means of a uniform magnetic field. In this Letter we report on the experimental observation of non-reciprocal effects in light scattering by the two dimensional lattice of magnetic vortices.

We begin with some phenomenological arguments in favor of non-reciprocal light scattering by the vortex particle. If one consider the light scattering cross-section summed over the polarization of incident and scattered light the reciprocity law takes simple form
\begin{equation} \label{Eq_1}
\sigma \left(\bf{k},\bf{k}',\bf{M}\left(\bf{r}\right)\right)=\sigma \left(-\bf{k}',-\bf{k},-\bf{M}\left(\bf{r}\right)\right),
\end{equation}
here $\sigma $ is the differential cross-section for the scattered light, $\bf{k}$ and $\bf{k}'$ are the wave vectors of the incident and scattered  beams, $\bf{M}\left(\bf{r}\right)$ is the magnetization spatial distribution. The term ``non-reciprocal effect'' implies one of the two equivalent  inequalities
\begin{equation} \label{Eq_2}
\sigma \left(\bf{k},\bf{k}',\bf{M}\left(\bf{r}\right)\right)\ne \sigma \left(-\bf{k}',-\bf{k},\bf{M}\left(\bf{r}\right)\right),
\end{equation}
\begin{equation} \label{Eq_3}
\sigma \left(\bf{k},\bf{k}',\bf{M}\left(\bf{r}\right)\right)\ne \sigma \left(\bf{k}',\bf{k},-\bf{M}\left(\bf{r}\right)\right).
\end{equation}

For the systems without a center of inversion the scattering cross-section can contain the term $\left(\left(\bf{k}+\bf{k}'\right)\cdot \bf{C}\right)$, where is $\bf C$ is a vector. This is linear in wavevector term which leads to non-reciprocal effects Eqs. (\ref{Eq_2},\ref{Eq_3}). According to reciprocity law for systems without spatial inversion the $\bf{C}$ vector should be a {\it polar vector} which also
change its sign under the time reversal. For a magnetic scatterer of centrosymmetrical shape made of a centrosymmetrical material, $\bf{C}$ can be chosen in the simplest form $\bf{C}=\alpha \left\langle \left[\bf{r}\times \bf{M}\left(\bf{r}\right)\right]\right\rangle $ which is a toroidal moment of the particle associated with the magnetic vorticity \cite{Pros08}, square brackets mean the spatial averaging over the scatterer, $\alpha$ is a constant. It follows from the above that the scattering of unpolarized light by the particle with the vortex magnetization distribution is non-reciprocal and has the contribution depending on the vorticity:
\begin{equation} \label{Eq_4}
\sigma \left(\bf{k},\bf{k}',\bf{M}\left(\bf{r}\right)\right)=...+\alpha \left(\left(\bf{k}+\bf{k}'\right)\cdot \left\langle \left[\bf{r}\times \bf{M}\left(\bf{r}\right)\right]\right\rangle \right).
\end{equation}

We carry out the experimental investigations of light scattering by lattices of the magnetic vortices in order to confirm the existence of such a contribution in the scattering cross-section.

\begin{figure}[t]
\includegraphics[width=3.25in, keepaspectratio=true]{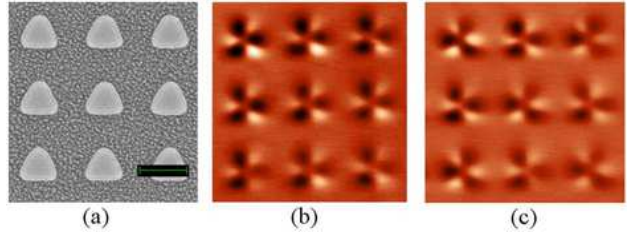}%
\caption{\label{Fig_1} (a) SEM image of the lattice of the cobalt triangles, the scale bar length is 1 $\mu$m, (b) MFM image of the remanent states after magnetizing along the base of the triangles. All magnetic vortices demonstrate the same direction of the vorticity. (c) MFM image of the remanent states after magnetizing along the height of the triangles. The vortices with both CW and CCW vortices are presented.}
\end{figure}

2D arrays of 30 nm thick polycrystalline triangular Co and CoFe dots is fabricated by electron beam lithography and lift-off technique on the surface of the amorphous SiO${}_{2}$ plate. The details of the technological procedures can be found in Ref.~\cite{Fra02}. The dots are arranged in a 400$\times$400 $\mu$m${}^{2}$ area and disposed in a square lattice with the period of 1.4 $\mu$m. The period makes it possible to observe diffraction maxima for the scattering of HeNe laser beam ($\lambda$=632 nm) used in our measurements. The size of the particles along a side is equal to 0.7 $\mu$m (Fig.~\ref{Fig_1}(a)).
This size is determined by two competing requirements: 1) to obtain the maximum possible volume of magnetic material, and 2) to avoid the significant interparticle magnetostatic interaction because it can affect the magnetization state of the particles \cite{Fra02}.

\begin{figure}[t]
\includegraphics[width=3.25in, keepaspectratio=true]{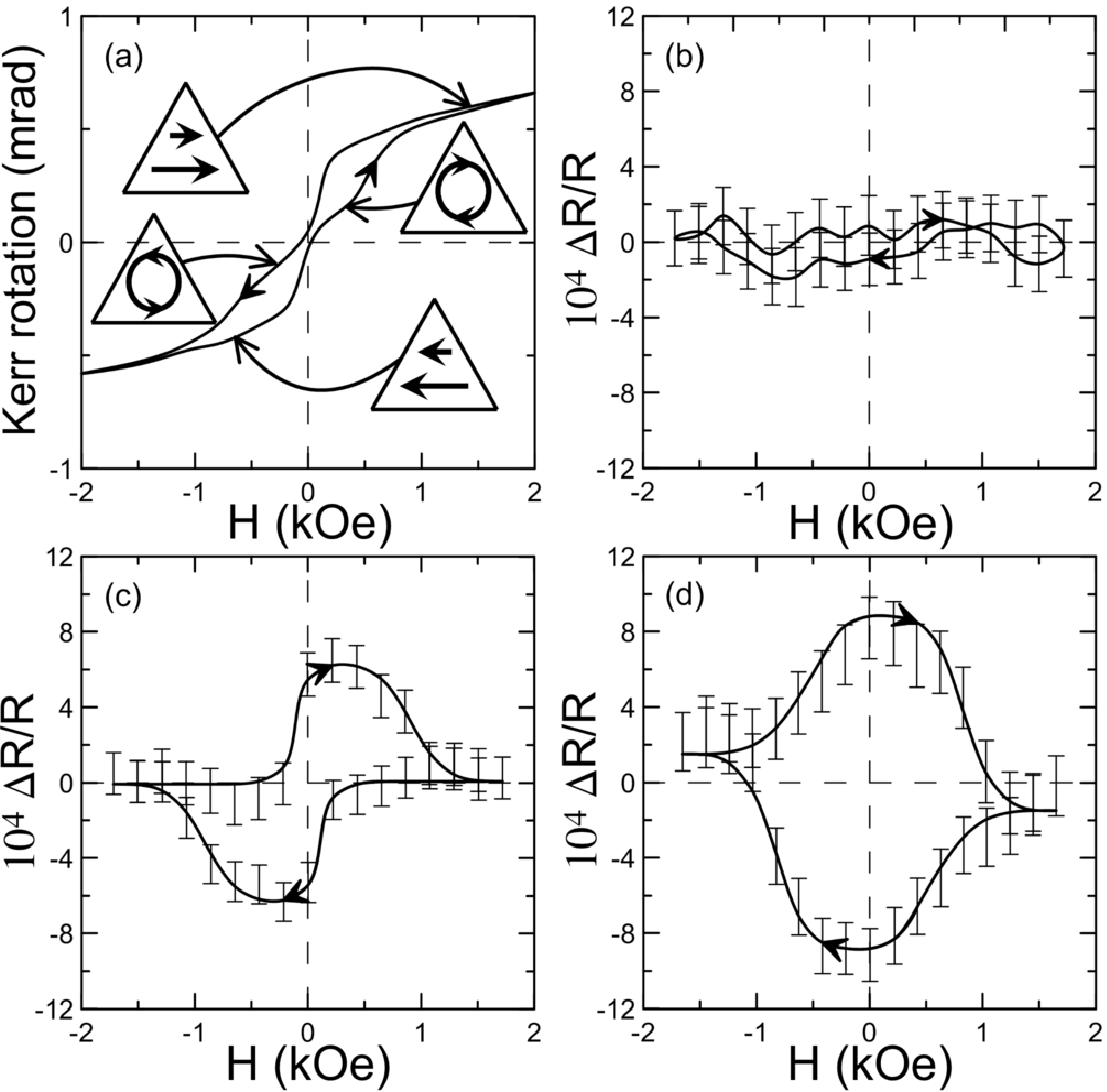}%
\caption{\label{Fig_2} (a) In-plane magnetization curve of the lattice of the ferromagnetic triangles measured by the magnetooptic Kerr rotation. Magnetic field is applied along the triangles side. (b-d) Relative change of the light intensity ($\Delta R=R(H)-\langle R(H)\rangle$, square brackets mean averaging over the magnetization cycle, $R(H)$ is the intensity of the diffracted light, $H$ is the magnitude of the external magnetic field) scattered in (-1,0)${}_{tr}$ maximum as a function of the applied magnetic field. The incident angle is 5${}^{\circ}$. (b) The field is directed along the triangles height. (c) The field is directed along the triangles side. Incident wave is s-polarized. (d) The field is directed along the triangles side. Incident wave is p-polarized. Error bars represent experimental data; solid lines are guides for eyes showing the direction of the hysteresis loop traversal during the magnetizing cycle. The distributions of the magnetization corresponding to hysteresis loop segments are represented schematically (view in z direction).}
\end{figure}

In zero external field the ground magnetic state of the triangle particle with the sizes mentioned above is the vortex state (Fig.~\ref{Fig_1}(b,c)). The left and right handed vortices have different sign of the vorticity and thus they should have different light scattering cross-section in accordance with Eq. (\ref{Eq_4}). The magnetization hysteresis loop of the lattice of the Co triangles for the case when the external field is applied along the triangle base has the shape usual for particles with the magnetic vortices (Fig.~\ref{Fig_2}(a)) \cite{Cow99}.
It is obtained by magneto-optical Kerr effect (MOKE) measurement in meridional configuration at room temperature. In high positive magnetic field all the particles are magnetized uniformly. When the magnetic field is decreased from saturation a magnetic vortex nucleates that is accompanied by an abrupt decrease in magnetization. If the field is directed along the base of the triangle to the right (Fig.~\ref{Fig_2}(c)) only counter clockwise (CCW) vortices enter the particles. This is the direct consequence of the non-centrosymmetric triangular shape of the particles. The magnetization states are verified with the magnetic force microscopy (MFM) investigations. Depending on the sample 90$\div$100\% of particles are in the same vortex state in zero field after such procedure (Fig.~\ref{Fig_1}(b)). In high negative field all the particles are uniformly magnetized in the field direction again. When magnetic field increases from high negative values the magnetization in the particles obtains clockwise (CW) vorticity. So by applying the uniform external field we can synchronically manipulate the vorticity of all the particles \cite{Yak10}. If the magnetic field is oriented along the triangle height the shape of the hysteresis loop is practically the same. But in this case the probability of the CW and CCW vortices nucleation is equal and there is a mixture of the vortices with both vorticities in the zero external field (Fig.~\ref{Fig_2}(c)).

\begin{figure}[t]
\includegraphics[width=3.25in, keepaspectratio=true]{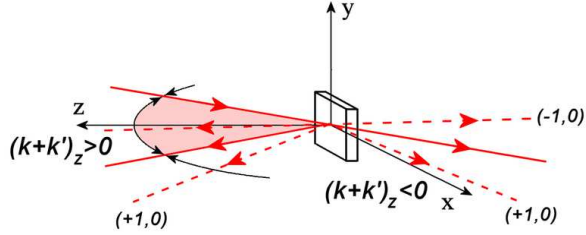}%
\caption{\label{Fig_3} The geometry of the experiment. Solid lines are incident, transmitted and specularly reflected beams, dashed lines represent diffracted beams. The grey-colored (red online) segment of the XZ plain corresponds to the directions with $(k+k')_{z}>0$.}
\end{figure}

The geometry of the optical measurements is depicted in Fig.~\ref{Fig_3}. The sample lies in the (x,y) plane with the lattice vectors oriented along x- and y- axes. Averaged toroidal moment $\left\langle \left[\bf{r}\times \bf{M}\left(\bf{r}\right)\right]\right\rangle $ is parallel to z-axis in this geometry. The sample is irradiated with the laser beam propagating in the (z,x) plane at the angles 5$\div$40${}^{\circ}$ with respect to to z-axis. Intensity of the scattered light is measured in four diffraction maxima lying in the same plane (dashed lines in Fig.~\ref{Fig_3}). These are ($\pm$1,0) maxima both for the transmitted and reflected light. The external magnetic field is directed in the plane of the sample parallel to x-axis. During the measurement, data are taken as a function of magnetic field to generate a hysteresis loop. The measurements are carried out separately for s- and p- polarization of the incident light. To compare the experimental results with the prediction Eq. (\ref{Eq_4}) the diffracted intensity for different polarization should be summed over.

To check the phenomenological predictions Eq. (\ref{Eq_4}) we investigate the dependence of the intensity of light scattered in the diffraction maxima on the direction of the magnetic particle vorticity. The measurements for different incident angles that allow exploring diffraction maxima with positive and negative values of $k_{z}+k'_{z}$ are carried out also.

The following main experimental optical results are summarized below:

({\it i}) During the magnetizing cycle (starting from the saturation field along the x-axis) the magnetic particles sequentially pass the following states: homogeneous (single-domain) $\to$ CCW vortex $\to$ homogeneous $\to$ CW vortex. Different magnitude of the intensity of light scattered by CW and CCW vortices leads to the appearance of the hysteresis loop (Fig.~\ref{Fig_2}(c,d)). Indeed in zero field we have different diffracted intensity for CW and CCW vortices. The shape of the hysteresis loop is different for s- and p- polarizations, but the direction of traversal of the hysteresis curve (i.e. sign of the effect) does not depend on the polarization. Thus, if one sums the diffracted intensity over the incident polarization the hysteresis loop does not disappear and one gets the polarization independent part of the effect (described by Eq. (\ref{Eq_4})).

({\it ii}) If the magnetizing field is directed along the height of the triangles no change in the intensity of the light scattered in the diffraction peaks is observed. Indeed, the CW and CCW vortices have the same probability to appear in this situation and consequently their number in the system is the same. This leads to the disappearance of the non-reciprocal effects.

\begin{figure}[t]
\includegraphics[width=3.25in, keepaspectratio=true]{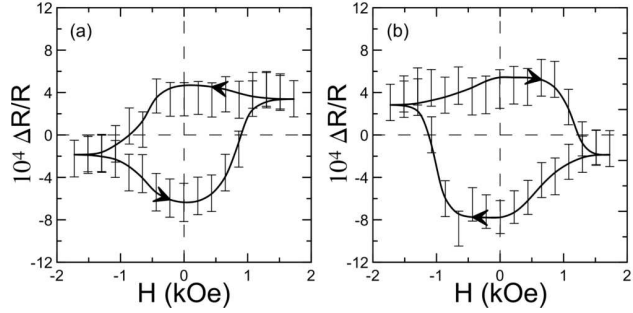}%
\caption{\label{Fig_4} Relative change of the intensity of the light scattered (a) in (-1,0)${}_{ref}$ maximum ($k_{z}+k'_{z}>0$) and (b) in (+1,0)${}_{ref}$ maximum ($k_{z}+k'_{z}<0$) measured simultaneously in a single run as a function of the applied magnetic field. The incident angle of the s-polarized light is 30${}^{\circ}$. Error bars represent experimental data, solid lines are guides for eyes showing the direction of the hysteresis loop traversal. The change of the hysteresis from counterclockwise type to clockwise type with the change of the sign of ($k_{z}+k'_{z}$) is evident.}
\end{figure}

({\it iii}) For the incident angle of 5${}^{\circ}$ the effect has the same sign for ($\pm$1,0) maxima both in reflection and transmission. This can be explained by the fact that in this geometry the value of $k_{z}+k'_{z}$ is always negative (see Fig.~\ref{Fig_3}). If the incident angle is equal to 30${}^{\circ}$ the sum $k_{z}+k'_{z}$ becomes positive for (-1,0)${}_{refl}$ diffraction maximum whereas it is still negative for (+1,0)${}_{refl}$ diffraction maximum (as in the previously described experiments with the incident angle of 5${}^{\circ}$). Simultaneous measurement of the intensity of light scattered in this diffraction maxima with the external magnetic field applied along the base of the triangle was carried out. We observe that the change in the intensity of the scattered light in these two maxima has different sign for the same direction of the vorticity. It is manifested in the opposite directions of the hysteresis loop circumvention in these cases (Fig.~\ref{Fig_4}).

So the experiments demonstrate that the intensity of the light diffracted by the vortex lattice depends on the scalar product $\left(\left(\bf{k}+\bf{k}'\right)\cdot \left\langle \left[\bf{r}\times \bf{M}\left(\bf{r}\right)\right]\right\rangle \right)$ i.e. has non-reciprocal character. Thus in spite of the zero average magnetization of the particles in the zero external field the existence of the magnetic vortex is manifested in the intensity of the diffracted light.

We propose two possible mechanisms of the appearance of the non-reciprocal term in the scattering cross-section. The first one is the excitation of an electric quadrupole moment in the particle with the vortex magnetization distribution under the impact of the uniform electric field of the incident wave. The quadrupole electric moment appears due to the anomalous Hall Effect and non-uniform magnetization distribution. The phase of quadrupole moment oscillation  depends on the vorticity of the particle. The interference of the waves radiated by quadrupole and dipole leads to the vorticity-dependent contribution to the scattering cross-section. The second possible mechanism is the excitation of electric dipole in the vortex particle under the influence of uniform magnetic field of the incident wave. It is well-known that magnetic field induces a vortex eddy current in the conductive particle. The influence of the magnetic vortex on this current leads to an additional contribution to the electrodipole moment of the particle. The sign of the contribution also depends on the vorticity.

As it can be seen from the Fig.~\ref{Fig_4} and Fig.~\ref{Fig_2} the intensity of the diffracted light is different for the saturation fields of opposite direction. This means that the term linear in magnetization $\alpha_{i,j}\left(k+k'\right)_{i}M_{j}$ appears in the scattering cross-section. Here $\alpha_{i,j}$ is a pseudotensor. The lattice of the triangles has only the reflection mirror plane (y,z) (here x is along the triangles base, y is along the triangles height, both x and y is along the lattice vectors, z is perpendicular to the sample surface). So the components of the pseudotensor $\alpha_{x,y}$, $\alpha_{x,z}$, $\alpha_{y,x}$, $\alpha_{z,x}$ are non zero. This is in agreement with the experimental data. From the Fig.~\ref{Fig_2}(b,d) one can see that the effect is zero when the magnetic field is directed along the y-axis ($\alpha_{z,y}=0$) and non zero when the field is along x-axis ($\alpha_{z,x}\ne0$). Also the effect change its sign when the z-projection of the vector $\bf{k}+\bf{k'}$ change the sign (see Fig.~\ref{Fig_4}).

To conclude, light diffraction by two-dimensional square lattice of triangle Co and CoFe nanoparticles in the vortex magnetic state is investigated. The peculiarity of the system is that all the particles have the same vorticity which can be manipulated by the uniform external magnetic field. We observed non-reciprocal intensity effect namely the dependence of the diffracted intensity on the particles vorticity. The observed effect could be described by the phenomenological expression $\alpha \left(\left(\bf{k}+\bf{k}'\right)\cdot \left\langle \left[\bf{r}\times \bf{M}\left(\bf{r}\right)\right]\right\rangle \right)$. Possible microscopic reasons of the effect are discussed also.

We are grateful to Dr. A.Yu. Aladyshkin for fruitful discussions.This work is supported by the Russian Foundation for Basic Research, the ``Dynasty'' foundation, OPTEC grant for young scientists, RF Agency for Education of Russian Federation (Rosobrazovanie), the Agency for Science and Innovation of Russian Federation, and the Common Research Center ``Physics and technology of micro- and nanostructures'' (contract No. 16.552.11.7007).

\bibliography{karashtin_text}

\end{document}